\documentclass[%
reprint,
superscriptaddress,
showpacs,
 amsmath,
 amssymb,
 aps,
]{revtex4-1}

\usepackage{graphicx}% Include figure files
\usepackage{latexsym}
\usepackage{dcolumn}% Align table columns on decimal point
\usepackage{bm}% bold math
\usepackage{epsfig}
\usepackage{graphics}
\usepackage{float}
\usepackage{hyperref}
\usepackage{braket}
\usepackage{bm,color}
\usepackage{amssymb}
\usepackage{amsmath}
\usepackage{amsfonts}
\usepackage{upgreek}
\usepackage{multirow}
\usepackage{hyperref}
\hypersetup{
colorlinks = true,
linkcolor = [rgb]{0.70,0.13,0.13},
citecolor = [rgb]{0.13,0.55,0.13},
urlcolor = [rgb]{0.25, 0.41, 0.88}}
\usepackage[mathlines]{lineno}% Enable numbering of text and display math
\newcommand{\ii}{\mathrm{i}}

\newcommand{\U}{\mathrm{U}}

\renewcommand{\O}{\mathrm{O}}

\newcommand{\dsE}{\mathbb{E}}

\newcommand{\dsR}{\mathbb{R}}
\newcommand{\dsC}{\mathbb{C}}

\newcommand{\scN}{\mathcal{N}}
\newcommand{\scL}{\mathcal{L}}

\newcommand{\scS}{\mathcal{S}}
\newcommand{\scO}{\mathcal{O}}

\newcommand{\scM}{\mathcal{M}}

\renewcommand{\Re}{\operatorname{Re}}
\renewcommand{\Im}{\operatorname{Im}}

\newcommand{\eq}[1]{\begin{equation}#1\end{equation}}
\newcommand{\eqs}[1]{\begin{equation}\begin{split}#1\end{split}\end{equation}}
\newcommand{\eqnref}[1]{Eq.\,\eqref{#1}}
\newcommand{\figref}[1]{Fig.\,\ref{#1}}

\newcommand{\appref}[1]{Supplementary Material \ref{#1}}
\newcommand{\refcite}[1]{Ref.\,\onlinecite{#1}}

\begin{document}

%\preprint{APS/123-QED}

\title{Machine Learning Holographic Mapping by Neural Network Renormalization Group}% Force line breaks with \\

\author{Hong-Ye Hu}
\affiliation{Department of Physics, University of California at San Diego, La Jolla, CA 92093, USA}
\author{Shuo-Hui Li}
\affiliation{Institute of Physics, Chinese Academy of Sciences, Beijing 100190, China}
\affiliation{University of Chinese Academy of Sciences, Beijing 100049, China}
\author{Lei Wang}
\affiliation{Institute of Physics, Chinese Academy of Sciences, Beijing 100190, China}
\affiliation{CAS Center for Excellence in Topological Quantum Computation, University of Chinese Academy of Sciences, Beijing 100190, China}
\affiliation{Songshan Lake Materials Laboratory, Dongguan, Guangdong 523808, China}
\author{Yi-Zhuang You}
\email{yzyou@physics.ucsd.edu}
\affiliation{Department of Physics, University of California at San Diego, La Jolla, CA 92093, USA}

\date{\today}
\begin{abstract}
The exact holographic mapping (EHM) provides an explicit duality map between a conformal field theory (CFT) configuration and a massive field propagating on an emergent classical geometry. However, designing the optimal holographic mapping is challenging. Here we introduce the neural network renormalization group as a universal approach to design generic EHM for interacting field theories. Given a field theory action, we train a flow-based hierarchical deep generative neural network to reproduce the boundary field ensemble from uncorrelated bulk field fluctuations. In this way, the neural network develops the optimal renormalization group transformations. Using the machine-designed EHM to map the CFT back to a bulk effective action, we determine the bulk geodesic distance from the residual mutual information. We apply this approach to the complex $\phi^4$ theory in two-dimensional Euclidian spacetime in its critical phase, and show that the emergent bulk geometry matches the three-dimensional hyperbolic geometry. 
\end{abstract}
\pacs{05.10.Cc, 11.25.Hf, 04.62.+v}% PACS, the Physics and Astronomy
\maketitle

%\tableofcontents

%\section{Introduction}

\section{Introduction}

The holographic duality, also known as the anti-de-Sitter space and conformal field theory correspondence (AdS/CFT)~\cite{Witten1998ASSH,Witten1998ASSTPTCGT,Gubser1998GTCFNST,Maldacena1999LLSFTS}, is a duality between a CFT  on a flat boundary and a gravitational theory in the AdS bulk with one higher dimension. It is intrinsically related to the renormalization group (RG) flow~\cite{de-Boer2000HRG,Skenderis2002LNHR,Heemskerk2011HWRG,Swingle2012CHSUER,Swingle2012ERH,Nozaki2012HGERQFT,Balasubramanian2013HIRG} of the boundary quantum field theory, since the dilation transformation, as a part of the conformal group, naturally corresponds to the coarse-graining procedure in the RG flow. The extra dimension emergent in the holographic bulk can be interpreted as the RG scale. In the traditional real-space RG~\cite{Kadanoff1966SLIMNT}, the coarse-graining procedure decimates irrelevant degrees of freedom along the RG flow, therefore the RG transformation is irreversible due to the information loss. However, if the decimated degrees of freedom are collected and hosted in the bulk, the RG transformation becomes a \emph{bijective} map between the degrees of freedom on the CFT boundary and the degrees of freedom in the AdS bulk. Such mappings, generated by information-preserving RG transforms, are called exact holographic mappings (EHM)~\cite{Qi2013EHMESG,Lee2015EHMFFS,Gu2016HDB2QAHS3TI}, which were first formulated for free fermion CFT. Similar idea was also implemented by multiscale entanglement renormalization ansatz (MERA)~\cite{Vidal2007ER,Evenbly2014CHEMSTES} as a hierarchical quantum circuit to simulate quantum state, as well as many of its generalizations~\cite{Haegeman2013ERQFRS,Lee2014QRGH,Mollabashi2014HGCQQFT,Leigh2014HGRGHSS,Lunts2015IH,Molina-Vilaplana2015IGERFQF,Miyaji2015BSHDTS,Wen2016HERTI,You2016EHMMLSSBRG, Cotler:2018ehb, Cotler:2018ufx}. Under the EHM, the boundary features of a quantum field theory of different scales are mapped to different depths in the bulk, and vice versa. The field variable deep in the bulk represents the overall or infrared (IR) feature, while the variable close to the boundary controls the detailed or ultraviolet (UV) feature. Such a hierarchical arrangement of information is often observed in deep neural networks, particularly in convolutional neural networks (CNN)~\cite{Goodfellow2016DL}. The similarity between renormalization group and deep learning has been discussed in several works~\cite{Beny2013DLRG,Mehta2014EMBVRGDL,Beny2015RGSI,Oprisa2017CDLMRG,Lin2017DDCLWW,Gan2017HDL}. Deep learning techniques have also been applied to construct the optimal RG transformations~\cite{Li2018NNRG,Koch-Janusz2018MINNRG} and to uncover the holographic geometry~\cite{You2018MLSGFEF,Hashimoto2018DLMC,Hashimoto2018DLH,Hashimoto2019ADBM}. 

In this work, we further explore the possibility of designing the EHM for interacting quantum field theories using deep learning approaches. We first point out that the information-preserving RG and deep generative model can be unified as the forward and backward application of the EHM, designing a good RG scheme is equivalent to training an optimal generative model to produce field configurations following the Boltzmann weight. We then propose that the information theoretical goal for a good EHM is to minimize the mutual information in the holographic bulk, which serves as a guiding principle for the machine to design RG rules. Bases on these understandings, we construct a flow-based hierarchical generative model~\cite{Dinh2016Density,Kingma2018Glow} with tractable and differentiable likelihood, which allows us to apply deep learning techniques to train the optimal EHM directly from the field theory action on the holographic boundary. We show that the fluctuation of neural network parameters corresponds to the gravitational fluctuation in the holographic bulk, and optimizing these parameters resembles searching for a classical geometry approximation. The machine-learned holographic mapping can be used to perform both the sampling task (mapping from bulk to boundary) and the inference task (mapping from boundary to bulk), providing us new tools to study both the boundary and the bulk theories. For the sampling task, we run the generative model to propose efficient global-update for boundary field configurations, which helps to boost the Monte Carlo simulation of the CFT. In the inference task, we push the boundary field theory to the bulk and establish the bulk effective theory, which enables us to probe the emergent dual geometry (on the classical level) by measuring the mutual information in bulk field.

\section{Renormalization Group and Generative Model}Renormalization group (RG) plays a central role in the study of quantum field theory (QFT) and many body physics. The RG transformation progressively coarse-grains the field configuration to extract relevant features. The coarse-graining rules (or the RG schemes) are generally model-dependent and requires human design. Take the real-space RG\cite{Kadanoff1966SLIMNT} for example: for a ferromagnetic Ising model, the RG rule should progressively extract the uniform spin components as the most relevant feature; however for an antiferromagnetic Ising model, the staggered spin components should be extracted instead; if the spin couplings are randomly distributed on the lattice, the RG rule can be more complicated. When it comes to the momentum-space RG\cite{Wilson1983RGCP}, the rule becomes to renormalize the low-energy degrees of freedom by integrating out the high-energy degrees of freedom. What is the general designing principle behind all these seemly different RG schemes? Can a machine learns to design the optimal RG scheme based on the model action?

\begin{figure}[htbp]
\begin{center}
\includegraphics[width=0.85\columnwidth]{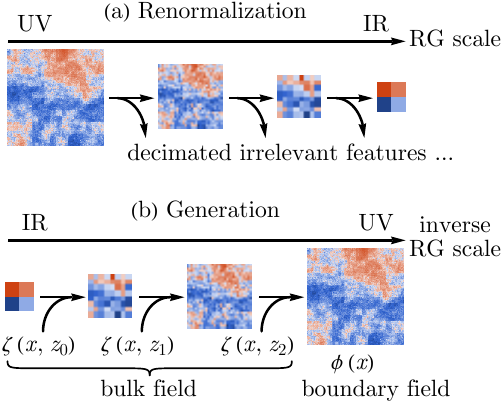}
\caption{Relation between (a) RG and (b) generative model. The inverse RG can be viewed as a generative model that generates the ensemble of field configurations from random sources. The random sources can are supplied at different RG scales (coordinated by $z$), which can be viewed as a field $\zeta(x,z)$ living in the holographic bulk with one more dimension. The original field $\phi(x)$ will be generated on the holographic boundary.}
\label{fig:RG}
\end{center}
\end{figure}

With these questions in mind, we take a closer look at the RG procedure in a lattice field theory setting. In the traditional RG approach, the RG transformation is invertible due to the information loss at each RG step when the irrelevant features are decimated, as illustrated in \figref{fig:RG}(a). However, if the decimated features are kept at each RG scale, the RG transformation can be inverted. Under the inverse RG flow, the decimated degrees of freedom $\zeta(x,z)$ are supplied to each layer (step) of the inverse RG transformation, such that the field configuration $\phi(x)$ can be regenerated, as shown in \figref{fig:RG}(b). Here we assume that the $\phi(x)$ field is defined in a flat Euclidean spacetime coordinated by $x=(x_1,x_2,\cdots)\in\dsR^d$, then $\zeta(x,z)$ will live on a manifold with one higher dimension, and the extra dimension $z$ corresponds to the RG scale. Given its close analogy to the holographic duality, we may view $\zeta(x,z)$ as the field in the holographic bulk and $\phi(x)$ as the field on the holographic boundary. The inverse RG can be considered as a deep generative model $G$, which organizes the bulk field $\zeta(x,z)$ to generate the boundary field $\phi(x)$, 
\eq{\phi(x)=G[\zeta(x,z)].}
The renormalization $G^{-1}$ and generation $G$ procedures are thus unified as the forward and backward maps of a bijective (invertible) map between the boundary and the bulk, known as the EHM.\cite{Qi2013EHMESG,Lee2015EHMFFS}

At the first glance, such an information-preserving RG does not seem to have much practical use, because it does not reduce the degrees of freedom and hence does not simplify our description. However, since the bulk field $\zeta(x,z)$ represents the irrelevant feature to be decimated under RG, it should look like independent random noise, which contains minimal amount of information. So instead of memorizing the bulk field configuration $\zeta(x,z)$ at each RG scale for reconstruction purpose, we can simply sample $\zeta(x,z)$ from uncorrelated (or weakly correlated) random source and serve them to the inverse RG transformation. Suppose the bulk field $\zeta(x,z)$ is drawn from a prior distribution $P_\text{prior}[\zeta]$, the transformation $\phi=G[\zeta]$ will deform the prior distribution to a posterior distribution $P_\text{post}[\phi]$ for the boundary field $\phi(x)$,
\eq{\label{eq:P_post}P_\text{post}[\phi]=P_\text{prior}[\zeta]\Big|\det\Big(\frac{\delta G[\zeta]}{\delta \zeta}\Big)\Big|^{-1},}
where $|\det(\delta_\zeta G)|^{-1}$ is the Jacobian determinant of transformation. In such manner, the objective of the inverse RG is not to reconstruct a particular original field configuration, but to generate an ensemble of field configurations $\phi(x)$, whose probability distribution $P_\text{post}[\phi]$ should better match the Boltzmann distribution
\eq{\label{eq:P_target}P_\text{target}[\phi]=e^{-S_\text{QFT}[\phi]}/Z_\text{QFT}} 
specified by the action functional $S_\text{QFT}[\phi(x)]$ of the boundary field theory, where $Z_\text{QFT}=\sum_{[\phi]}e^{-S_\text{QFT}[\phi]}$ denotes the partition function.

This setup provides us a theoretical framework to discuss the designing principles of a good RG scheme. We propose two objectives for a good RG scheme (or EHM): the RG transformation should aim at decimating irrelevant features and preserving relevant features, and the inverse RG must aim at generating field configurations matching the target field theory distribution $P_\text{target}[\phi]$ in \eqnref{eq:P_target}. An information theoretic criterion for  ``irrelevant'' features is that they should have minimal mutual information, so the prior distribution $P_\text{prior}[\zeta]$ should be chosen to minimize the mutual information between bulk fields at different points, i.e. $\min I(\zeta(x,z):\zeta(x',z'))$. We will refer to this designing principle as the minimal bulk mutual information (minBMI) principle, which is a general information theoretic principle behind different RG schemes and is independent of the notion of field pattern or energy scale. The close relation between RG and deep learning has been thoroughly discussed in several early works\cite{Mehta2014EMBVRGDL,Oprisa2017CDLMRG,Lin2017DDCLWW}. However, as pointed out in \refcite{Koch-Janusz2018MINNRG,Lenggenhager2018ORGTFIT}, the hierarchical architecture itself can not guarantee the emergence of RG transformation in a deep neural network. Additional information theoretic principles must be imposed to guild the learning. In light of this observation, \refcite{Koch-Janusz2018MINNRG,Lenggenhager2018ORGTFIT} proposed the maximal real-space mutual information (maxRSMI) principle, which aims at maximizing the mutual information between the coarse-grained field and the fine-grained field in the surrounding environment. Our minBMI principle is consistent with and more general than the maxRSMI principle (see \appref{sec:minBMI} for detailed discussion about the relation between these two principles).

In the simplest setting, we can hard code the minBMI principle by assigning the prior distribution to the uncorrelated Gaussian distribution,
\eq{\label{eq:P_prior1}P_\text{prior}[\zeta]=\scN[\zeta;0,1]\propto e^{-\|\zeta\|^2},} 
where $\|\zeta\|^2=\sum_{x,z}|\zeta(x,z)|^2$. Hence the mutual information vanishes for every pair of points in the holographic bulk. Given the prior distribution, the problem of finding the optimal EHM boils down to training the optimal generative model $G$ to minimize the Kullback-Leibler (KL) divergence between the posterior distribution $P_\text{post}[\phi]$ in \eqnref{eq:P_post} and the target distribution $P_\text{target}[\phi]$ in \eqnref{eq:P_target}, i.e. $\min\scL$ with
\eqs{\label{eq:L}\scL&=\mathsf{KL}(P_\text{post}[\phi]\parallel P_\text{target}[\phi])\\
&=\mathop{\dsE}\limits_{\zeta\sim P_\text{prior}}S_\text{QFT}[G[\zeta]]+\ln P_\text{prior}[\zeta]-\ln\det\Big(\frac{\delta G[\zeta]}{\delta \zeta}\Big),}
where $\dsE_{\zeta\sim P_\text{prior}}$ denotes the average over the ensemble of $\zeta$ drawn from the prior distribution. This fits perfectly to the framework of flow-based generative models\cite{Dinh2016Density,Kingma2018Glow} in machine learning, which can be trained efficiently thanks to its tractable and differentiable posterior likelihood. We model the bijective map $G$ by a neural network (to be detailed later) with trainable network parameters. We initiate the sampling from the bulk $\zeta\sim P_\text{prior}$ and push the bulk field to the boundary by $\phi=G[\zeta]$, collecting the logarithm of the Jacobian determinant along the way. Given the action $S_\text{QFT}[\phi]$, we can evaluate the loss function $\scL$ in \eqnref{eq:L} and back propagate its gradient with respect to the network parameters. We then update the network parameters by stochastic gradient descent. We iterate the above steps to train the neural network. In this way, simply by presenting the QFT action $S_\text{QFT}$ to the machine, the machine learns to design the optimal RG transformation $G$ by keep probing $S_\text{QFT}$ with various machine generated field configurations. Thus our algorithm may be called the neural network renormalization group (neural RG)\cite{Li2018NNRG}, which can be implemented using the deep learning platforms such as TensorFlow\cite{MartinAbadi2015TLMLHS}.

\section{Holographic Duality and Classical Approximation}
We would like to provide an alternative interpretation of the loss function $\scL$ in \eqnref{eq:L} in the context of holographic duality, which will deepen our understanding of the capabilities and limitations of our approach. Suppose we can sample the boundary field configuration $\phi(x)$ from the target distribution $P_\text{target}[\phi]$ and map $\phi(x)$ to the bulk by apply the EHM along the RG direction $\zeta=G^{-1}[\phi]$, the obtained bulk field $\zeta(x,z)$ will follow the distribution 
\eqs{P_\text{bulk}[\zeta]&=P_\text{target}[\phi]\det(\delta_\phi G^{-1}[\phi])^{-1}\\
&=Z_\text{QFT}^{-1}e^{-S_\text{QFT}[G[\zeta]]}\det(\delta_\zeta G).}
Then the normalization of the bulk field probability distribution $\sum_{[\zeta]}P_\text{bulk}[\zeta]=1$ implies that the QFT partition function $Z_\text{QFT}=\sum_{[\phi]}e^{-S[\phi]}$, which was originally defined on the holographic boundary, can now be written in terms of the bulk field $\zeta$ as well
\eq{\label{eq:Z_QFT dual}Z_\text{QFT}=\sum_{[\zeta]}e^{-S_\text{QFT}[G[\zeta]]+\ln\det(\delta_\zeta G)}.}
Note that $Z_\text{QFT}$ is by definition independent of $G$, we are allowed to sum over all possible $G$ on both sides of \eqnref{eq:Z_QFT dual}, which establishes a duality between the following two partition functions
\eq{\label{eq:duality}Z_\text{QFT}=\sum_{[\phi]}e^{-S_\text{QFT}[\phi]}\leftrightarrow Z_\text{grav}=\sum_{[\zeta,G]}e^{-S_\text{grav}[\zeta,G]},}
with the bulk theory $S_\text{grav}$ given by
\eq{\label{eq:S_grav}S_\text{grav}[\zeta,G]=S_\text{QFT}[G[\zeta]]-\ln\det(\delta_\zeta G).}
By ``duality'' we mean that $Z_\text{QFT}$ and $Z_\text{grav}$ only differ by a proportionality constant (as $Z_\text{grav}=\sum_{[G]}Z_\text{QFT}$), so they are equivalent descriptions of the same physics theory. $S_\text{grav}[\zeta,G]$ describes how the bulk variables $\zeta$ (matter field) and the neural network $G$ (geometry) would fluctuate and interact with each other, which resembles a ``quantum gravity'' theory in the holographic bulk. The bulk has more degrees of freedom than the boundary, as there can be many different choices of $\zeta$ and $G$ that leads to the same boundary field configuration $\phi=G[\zeta]$. This is a gauge redundancy in the bulk theory, which covers the diffeomorphism invariance as well as the interchangeable role between matter and spacetime geometry in a gravity theory. At this level, the bulk theory looks intrinsically nonlocal and the geometry can fluctuate strongly.

However, it is usually more desired to work with quantum gravity theories with a classical limit, which describe weak fluctuations (matter fields and gravitons) around a classical geometry. Although not every CFT admits a classical gravity dual, we still attempt to find the \emph{classical approximation} of the dual quantum gravity theory, neglecting the fluctuation of $G$. Such classical approximations could serve as a starting point on which gravitational fluctuations may be further investigated in future works. Aiming at a classical geometry, we look for the optimal $G$ that maximizes its marginal probability $P_\text{EHM}[G]=Z_\text{grav}^{-1}\sum_{[\zeta]}e^{-S_\text{grav}[\zeta,G]}$ with the bulk matter field $\zeta$ traced out. This optimization problem seems trivial, because according to \eqnref{eq:Z_QFT dual}, $P_\text{EHM}[G]=Z_\text{QFT}/Z_\text{grav}$ is independent of $G$. It is understandable that any choice of $G$ is equally likely if we have no preference on the prior distribution $P_\text{prior}[\zeta]$ of the bulk matter field, because there is a trade-off between $G$ and $P_\text{prior}$ that one can always adjust $P_\text{prior}$ to compensate the change in $G$. Such a trade-off behavior is fundamentally required by the gauge redundancy in the bulk gravity theory. To fix the gauge, we evoke the minBMI principle to bias the bulk matter field towards independent random noise, such that the classical solution of $G$ will look like a RG transformation, in line with our expectation for a holographic mapping. Choosing a minBMI prior distribution such as \eqnref{eq:P_prior1}, $P_\text{EHM}[G]$ can be cast into
\eq{\label{eq:P_EHM}P_\text{EHM}[G]=\mathop{\dsE}\limits_{\zeta\sim P_\text{prior}}\frac{P_\text{target}[G[\zeta]]}{P_\text{post}[G[\zeta]]}\geq e^{-\scL},}
which is bounded by $e^{-\scL}$ from below, with $\scL$ being the KL divergence between $P_\text{post}$ and $P_\text{target}$ as defined in \eqnref{eq:L}. Therefore the objective of maximizing $P_\text{EHM}[G]$ can be approximately replaced by minimizing the loss function $\scL$, which is no longer a trivial optimization problem. From this perspective, the loss function $\scL$ can be approximately interpreted as the action (negative log-likelihood) for the holographic bulk geometry associated to the EHM $G$. Minimizing the loss function corresponds to finding the classical saddle point solution of the bulk geometry. We will build a flow-based generative model to parameterize $G$ and train the neural network using deep learning approaches. The fluctuation of neural network parameters in the learning dynamics reflects (at least partially) the gravitational fluctuation in the holographic bulk. 

At the classical saddle point $G_*=\text{argmin}_G\scL$, we may extract an effective theory for the bulk matter field
\eqs{\label{eq:S_eff}S_\text{eff}[\zeta]&\equiv S_\text{grav}[\zeta,G_*]\\
&=\|\zeta\|^2+\ln P_\text{post}[G_*[\zeta]]-\ln P_\text{target}[G_*[\zeta]].}
As the KL divergence $\scL=\mathsf{KL}(P_\text{post}\parallel P_\text{target})$ is minimized after training, we expect $P_\text{post}$ and $P_\text{target}$ to be similar, such that their log-likelihood difference $\ln P_\text{post}-\ln P_\text{target}$ will be small, so the effective theory $S_\text{eff}[\zeta]$ will be dominated by the first term $\|\zeta\|^2$ in \eqnref{eq:S_eff}, implying that the bulk field $\zeta$ will be massive. The small log-likelihood difference further  provides kinetic terms (and interactions) for the bulk field $\zeta$, allowing it to propagate on a classical background that is implicitly specified by $G_*$. In this way, the bulk field will be correlated in general. Even though one of our objectives is to minimize the bulk mutual information as much as possible, the machine-learned EHM typically cannot resolve all correlations in the original QFT, so the residual correlations will be left in the bulk field $\zeta$ as described by the log-likelihood difference in \eqnref{eq:S_eff}. The mismatch between $P_\text{post}$ and $P_\text{target}$ may arise from several reasons: first, limited by the design of the neural network, the generative model $G$ may not be expressive enough to precisely deform the prior distribution to the target distribution; second, even if $G$ has the sufficient representation power, the training may not be able to converge to the global minimum; finally and perhaps the most fundamental reason is not every QFT has a classical gravitational dual, the bulk theory should be quantum gravity in general. Taking the classical approximation and ignoring the gravitational fluctuation leads the unresolvable correlation and interaction for the matter field $\zeta$ that has to be kept in the bulk. 

Nevertheless, our framework could in principle include fluctuations of $G$ by falling back to $Z_\text{grav}$ in \eqnref{eq:duality}. We can either model the marginal distribution $P_\text{EHM}[G]$ by techniques like graph generative models, or directly analyze the gravitational fluctuations by observing the fluctuations of neural network parameters in the learning dynamics as mentioned below \eqnref{eq:P_EHM}. We will leave these ideas for future exploration. In the following, we will use a concrete example, a 2D compact boson CFT on a lattice, to illustrate our approach of learning the EHM as a generative model and to demonstrate its applications in both the sampling and the inference tasks.

\section{Application to Complex $\phi^4$ Model}
We consider a lattice field theory defined on a 2D square lattice, described by the Euclidean action
\eq{S_\text{QFT}[\phi]=-t\sum_{\langle ij\rangle}\phi_i^*\phi_j+\sum_{i}(\mu|\phi_i|^2+\lambda|\phi_i|^4),}
where $\phi_i \in\dsC$ is a complex scalar field defined on each site $i$ of a square lattice and $\langle ij\rangle$ denotes the summation over all nearest neighbor sites. The model has a global $\U(1)$ symmetry, under which the field rotates by $\phi_i\to e^{\ii\varphi}\phi_i$ on every site. We choose $\mu=-200+2t$ and $\lambda=25$ to create a deep Mexican hat potential that basically pins the complex field on a circle $\phi_i=\sqrt{\rho}e^{\ii\theta_i}$ of radius $\sqrt{\rho}=2$. In this way, the field theory falls back to the XY-model $S_\text{QFT}=-\tfrac{1}{T}\sum_{\langle ij \rangle}\cos(\theta_i-\theta_j)$ with an effective temperature $T=(\rho t)^{-1}$. By tuning the temperature $T$, the model exhibits two phases: the low-$T$ algebraic liquid phase with a power-law correlation $\langle\phi_i^*\phi_j\rangle\sim|x_i-x_j|^{-\alpha}$ and the high-$T$ disordered phase with a short-range correlation. The two phases are separated by the Kosterlitz-Thouless (KT) transition. Several recent works\cite{Beach2018MLVKT,Zhang2018MLPTPM,Rodriguez-Nieva2018ITOUML,Zhou2018RGNNSFT} have focused on applying machine learning method to identify phase transitions or topological defects (vortices). Our purpose is different here: we stay in the algebraic liquid phase, described by a Luttinger liquid CFT, and seek to develop the optimal holographic mapping for the CFT.

\begin{figure}[htbp]
\begin{center}
\includegraphics[width=0.94\columnwidth]{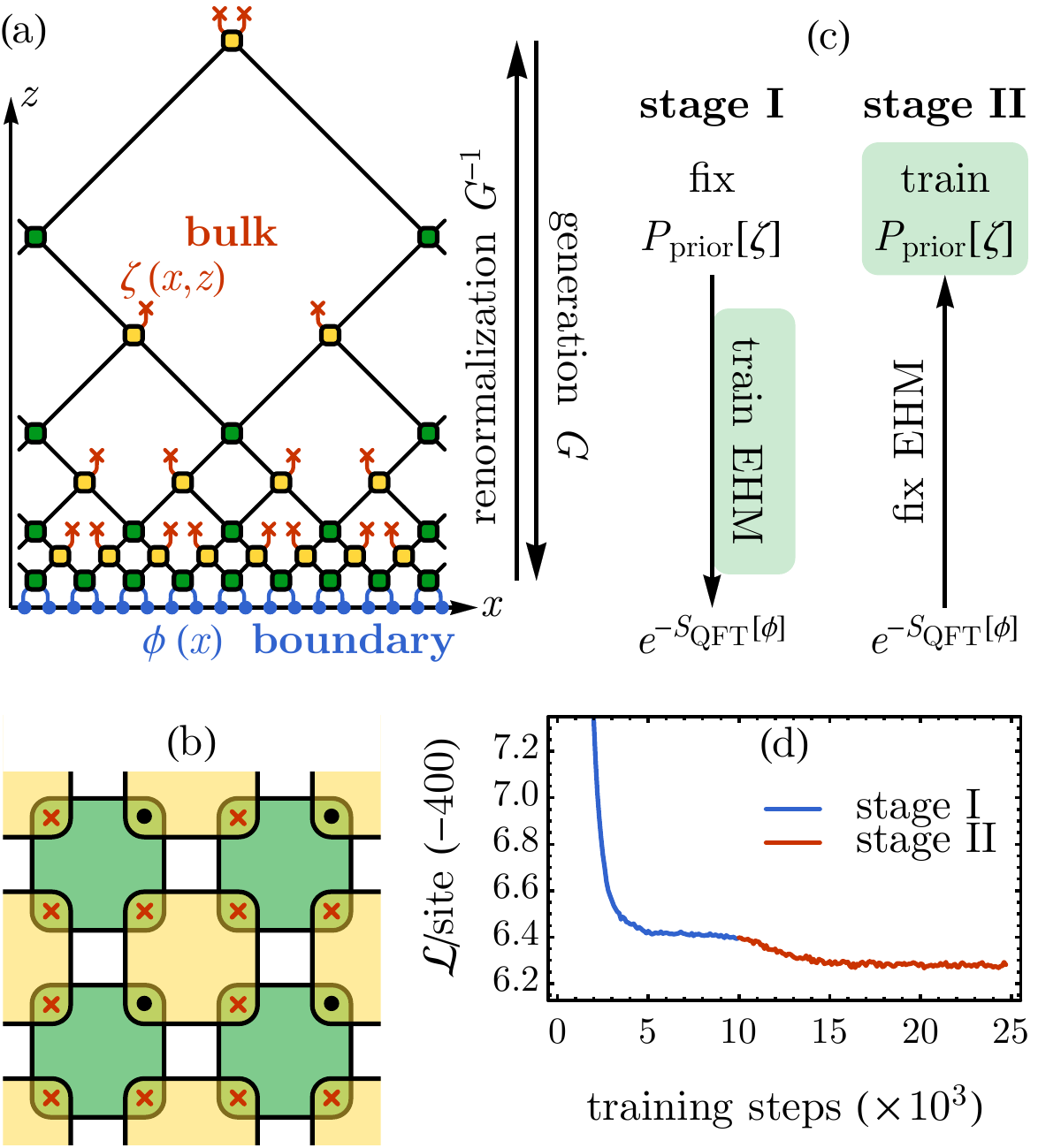}
\caption{(a) Side view of the neural-RG network. $x$ is the spatial dimension(s) and $z$ corresponds to the RG scale. There are two types of blocks: disentanglers (dark green) and  decimators (light yellow). The network forms an EHM between the boundary variables (blue dots) and the bulk variables (red crosses). (b) Top view of one RG layer in the network. Disentanglers and decimators interweave in the spacetime (taking two-dimensional spacetime for example). Each decimator pushes the coarse-grained variable (black dot) to the higher layer and leaves the decimated variables (red crosses) in the holographic bulk. (c) The training contains two stages. In the first stage, we fix the prior distribution $P[\zeta]$ to be uncorrelated Gaussian and train the EHM $G$ to bring it to the Boltzmann distribution of the CFT. In the second stage, we learn the prior distribution with the trained EHM held fixed. (d) The behavior of the loss function $\scL$ in the two training stages.}
\label{fig:network}
\end{center}
\end{figure}

\begin{figure}[htb]
\begin{center}
\includegraphics[width=0.55\columnwidth]{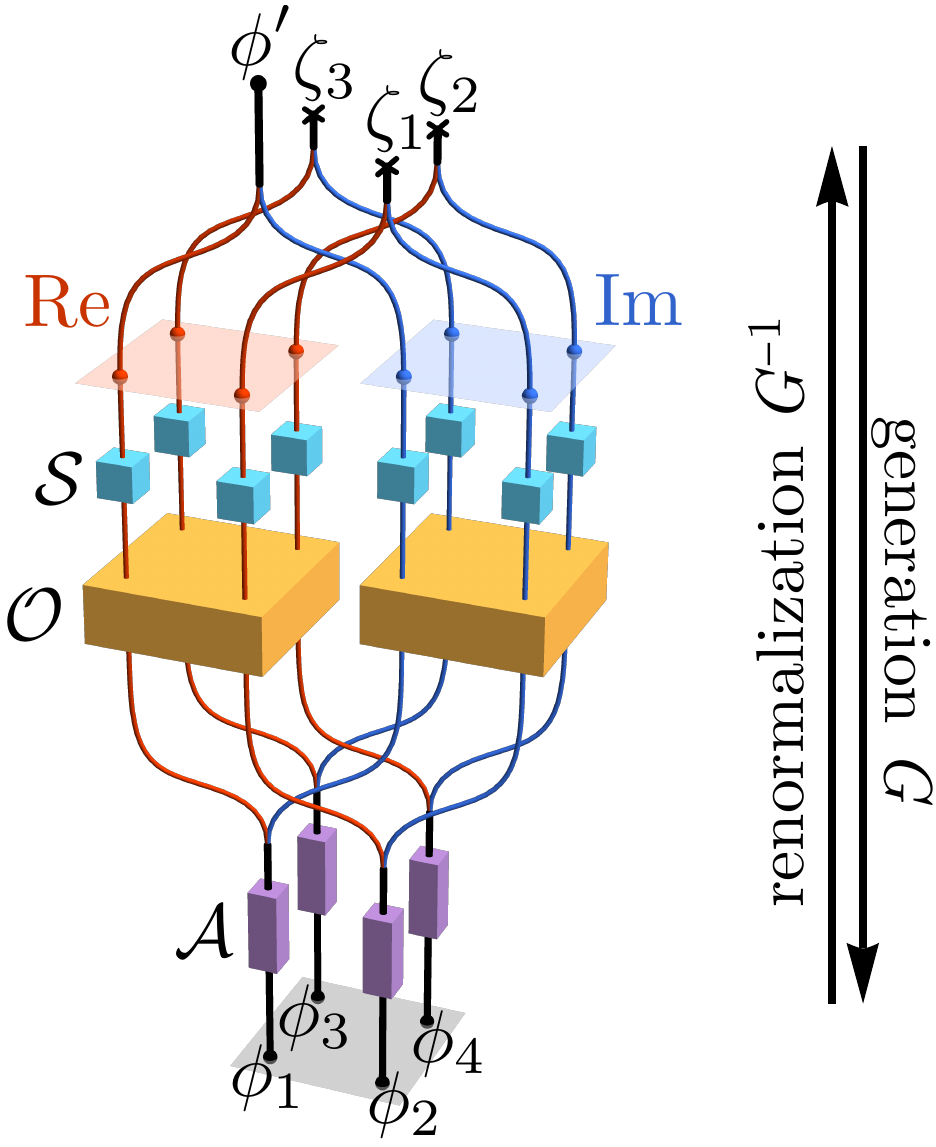}
\caption{Neural network architecture within a decimator block (the disentangler block shares the same architecture). Starting from the renormalized variable $\phi'$ and the bulk noise $\zeta_{1,2,3}$ as complex variables, the $\Re$ and $\Im$ channels are first separated, then $\scS$ applies the scaling separately to the four variables within each channel and $\scO$ implements the $\O(4)$ transformation that mixes the four variables together. $\scS$ and $\scO$ are identical for $\Re$ and $\Im$ channels to preserve the $\U(1)$ symmetry. Then the channels merge into complex variables followed by element wise non-linear activation describe by an invertible $\U(1)$-symmetric map $\phi_i\mapsto (\phi_i/|\phi_i|)\sinh|\phi_i|$.}
\label{fig:bijector}
\end{center}
\end{figure}

We design the generative model $G$ as a bijective deep neural network following the architecture of the neural network renormalization group (neural-RG) proposed by \refcite{Li2018NNRG}. Its structure resembles the MERA network~\cite{Vidal2007ER} as depicted in \figref{fig:network}(a). Each RG step contains a layer of disentangler blocks (like CNN convolutional layer) to resolve local correlations, and a layer of decimator blocks (like CNN pooling layer) to separate the renormalized and decimated variables. Given that the spacetime dimension is two on the boundary, we can overlay decimators on top of disentanglers in an interweaving manner as shwon in \figref{fig:network}(b). Both the disentangler and the decimator are made of three bijective layers: a linear scaling layer, an orthogonal transformation layer and an invertible non-linear activation layer, as arranged in \figref{fig:bijector} (see \appref{sec:bijectors} for more details). They are designed to be invertible, non-linear and $\U(1)$-symmetric transformations, which are used to model generic RG transformations for the complex $\phi^4$ model. The bijector parameters are subject to training (training procesure and number of parametes are specified in \appref{sec:NNtraining}). The Jacobian matrix of these transformations are calculable.  After each decimator, only one renormalized variable flows to the next RG layer, and the other three decimated variables are positioned into the bulk as little crosses as shown in \figref{fig:network}(a) and (b). The entire network constitutes an EHM between the original boundary field $\phi(x)$ and the dual field $\zeta(x,z)$ in the holographic bulk. 

We start with a $32\times32$ square lattice as the holographic boundary and build up the neural-RG network. The network will have five layers in total. Since the boundary field theory has a gobal $\U(1)$ symmetry, the bijectors in the neural network are designed to respect the $\U(1)$ symmetry (see \figref{fig:bijector}), such that the bulk filed also preserves the $\U(1)$ symmetry. The training will be divided into two stages, as pictured in \figref{fig:network}(c). In the training stage I, we fix the prior distribution in \eqnref{eq:P_prior1} and train the network parameters in the generative model $G$ to minimize the loss function $\scL$. The training method is outlined below \eqnref{eq:L}. The loss function $\scL$ decays with training steps, whose typical behavior is shown in \figref{fig:network}(d). We will discuss the stage II training later.

\begin{figure}[htbp]
\centering
\includegraphics[width = 0.95\linewidth]{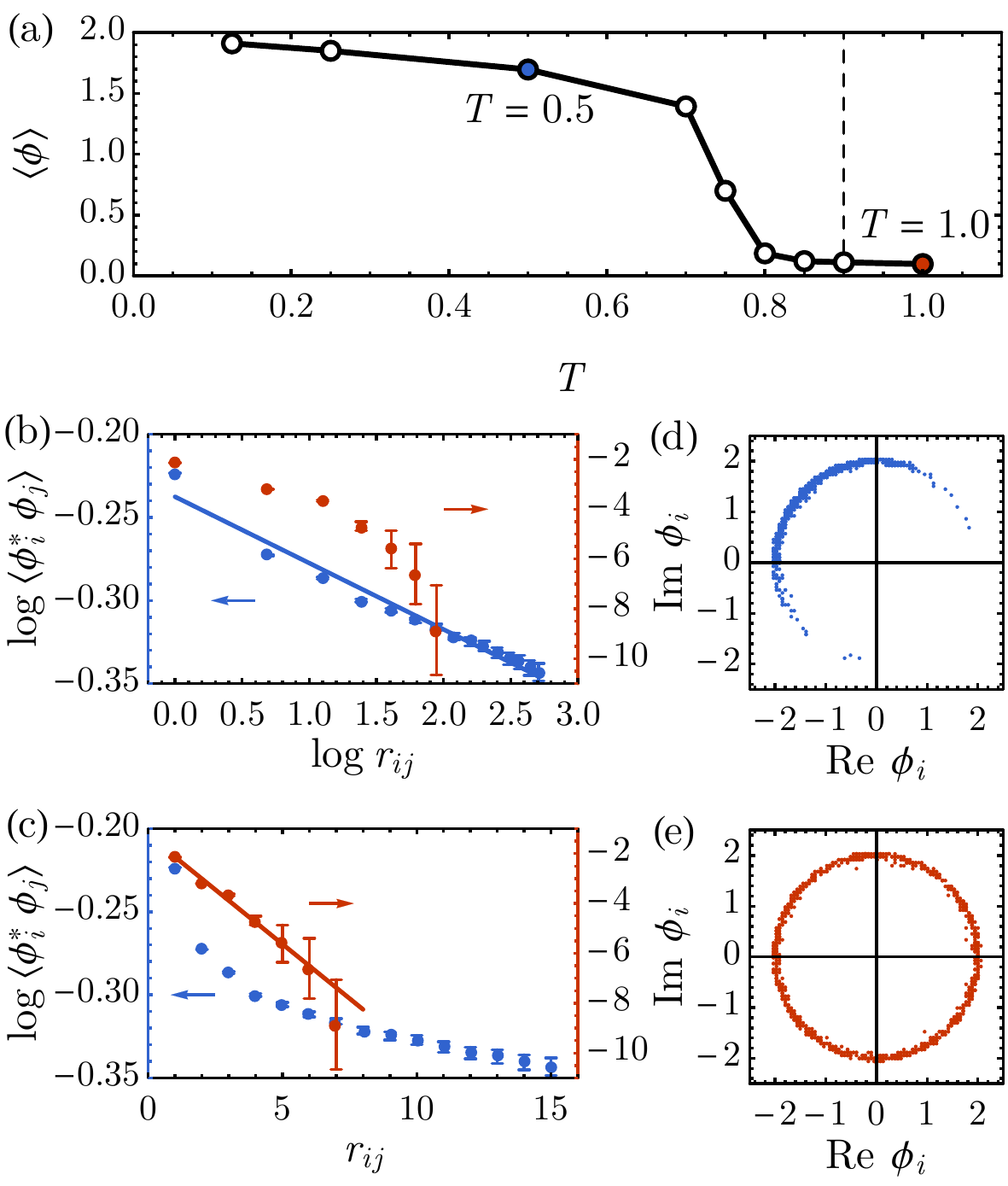}
\caption{Performance of the trained EHM for the complex $\phi^4$ theory. (a) Order parameter $\langle\phi\rangle$ v.s. temperature $T$. Different models are trained separately at different temperature. For finite-sized system, $\langle\phi\rangle$ crosses over to zero around the KT transition. Correlation function $\langle\phi_i^*\phi_j\rangle$ scaling in log-log plot (b) and log-linear plot (c). Distribution of $\phi_i$ in a single sample generated by the neural network trained in (d) the algebraic liquid phase and (e) the disordered phase.
\label{fig:phi4} }
\end{figure}

We perform the stage I training for several neural networks at different temperatures $T$ separately, i.e.~we use $S_\text{QFT}[\phi]$ of different parameters to train different neural networks. After training, each neural network can generate configurations of the boundary field $\phi$ from the bulk uncorrelated Gaussian field $\zeta$ efficiently.  To test how well these generative models work, we measured the order parameter $\langle\phi\rangle$ and the correlation function $\langle \phi_i^*\phi_j\rangle$ using the field configurations generated by the neural network. Although the order parameter $\langle\phi\rangle$ is expected to vanish in the thermodynamic limit, for our finite-size system, it is not vanishing  and can exhibit a crossover around the KT transition, as shown in \figref{fig:phi4}(a). The cross over temperature $T\simeq 0.9$ agrees with the previous Monte Carlo study~\cite{Olsson1995MCATMICWKRE,Hasenbusch1997CRTISMBMM,Hasenbusch2005TMTTHMCS} of the KT transition temperature $T_\text{KT}=0.8929$ in the two-dimensional XY model. We measure the correlation function $\langle \phi_i^*\phi_j\rangle$ at two different temperatures: one at $T=0.5$ in the algebraic liquid phase, one at $T=1.0$ in the disordered phase. We plot the two-point function $\langle \phi_i^*\phi_j\rangle$ as a function of the Euclidean distance $r_{ij}\equiv |x_i-x_j|$ (on the square lattice) in both the log-log scale as \figref{fig:phi4}(b) and the log-linear scale as \figref{fig:phi4}(c). The comparison shows that the correlation function in the algebraic liquid (or the disordered) phase fits better to the power-law (or the exponential) decay. \figref{fig:phi4}(d) shows the statistics of $\phi_i$ in one sample generated by the machine trained in the algebraic liquid phase. It exhibits the ``spontaneous symmetry breaking'' behavior due to the finite-size effect, although accumulating over multiple samples will restore the $\U(1)$ symmetry. However, similar plot \figref{fig:phi4}(e) in the disordered phase respects the $\U(1)$ symmetry in every single sample. Based on these tests, we can conclude that the neural network has learned to generate field configurations $\phi(x)$ that reproduce the correct physics of the complex $\phi^4$ model. The trained generative model $G$ maps an almost uncorrelated bulk field $\zeta$ to a correlated boundary field $\phi$, and vice versa, therefore $G$ provides a good EHM for the $\phi^4$ theory. 

\begin{figure}[thb]
\begin{center}
\includegraphics[width=\columnwidth]{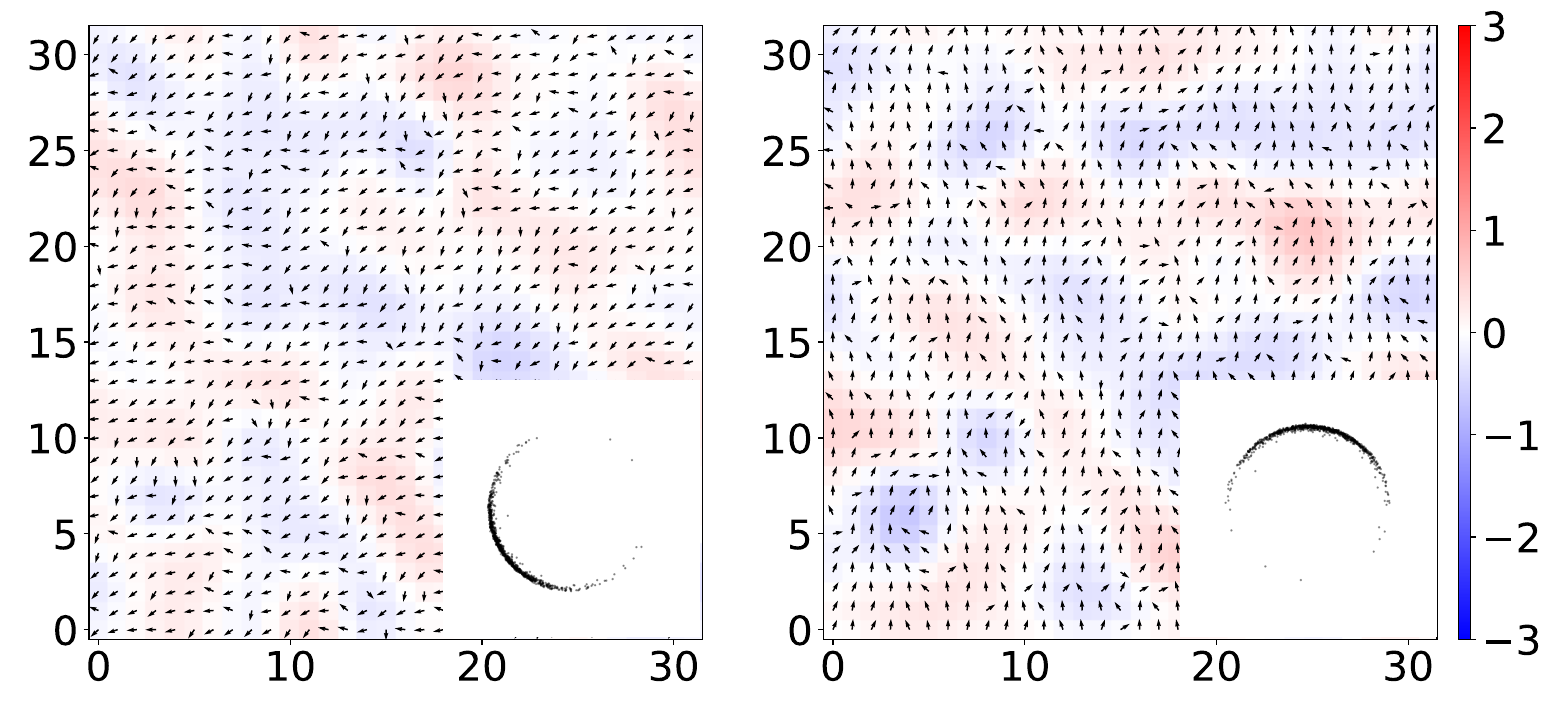}
\caption{The boundary field configuration $\phi$ before (left) and after (right) a local update in the most IR layer of the bulk field $\zeta$. The complex field $\phi_i$ is represented by the small arrow on each site. The background color represents the vorticity. The inset shows the distribution of $\phi_i$ in the complex plane.}
\label{fig:local_global}
\end{center}
\end{figure}

The machine-learnt EHM can be useful in both the backward and forward directions. The backward mapping from bulk to boundary provides efficient sampling of the CFT configurations, which can be used to boost the Monte Carlo simulation of the CFT. The forward mapping from boundary to bulk enables direct inference of bulk field configurations, allowing us to study the bulk effective theory and to probe the bulk geometry. Let us first discuss the sampling task. The EHM establishes a mapping between the massive bulk field $\zeta$ and the massless boundary field $\phi$. The bulk field admits efficient sampling in terms of local update, because the it is uncorrelated (or short-range correlated). Local updates in the bulk gets maps to global updates on the boundary, which allows us to sample the critical boundary field efficiently, minimizing the effect of critical slowdown. To demonstrate this, we tweak the bulk field in the most IR layer. Under the EHM, we observe a global change of the boundary field configuration as shown in \figref{fig:local_global}. It is interesting to note that the change of the IR bulk field basically induces a global $\U(1)$ rotation of $\phi_i$ (see the insets of \figref{fig:local_global}), which corresponds the ``Goldstone mode'' associated to the ``spontaneous symmetry breaking'' in the fine-sized system, showing that the machine can identify the order parameter as the relevant IR degrees of freedom without prior knowledge about low-energy modes of the system. We also check that the Hamiltonian Monte Carlo sampling in the bulk converges much faster compared to applying the same algorithm on the boundary (see \appref{sec:MC} for more evidences). In connection to several recent works, our neural-RG architecture can be integrated to self-learning Monte Carlo approaches\cite{Liu2016SMCMFS,Aoki2016RBMLRIM,Huang2017AMCSWRBM,Liu2017SMCM,Nagai2017SMCMCA,Tanaka2017TRAML,Nagai2018SMCMWBNN} to boost the numerical efficiency in simulating CFTs. The inverse RG transformation can also be used to generate super-resolution samples\cite{Efthymiou2018SIMWCNN} for finite-size extrapolation of thermodynamic observables.

Now let us turn to the inference task. We can use the optimal EHM to push the boundary field back into the bulk and investigate the effective bulk theory $S_\text{eff}[\zeta]$ induced by the boundary CFT. As analyzed below \eqnref{eq:S_eff}, the mismatch between $P_\text{post}$ and $P_\text{target}$ will give rise to the residual correlation (mutual information) of the bulk matter field, which can be used to probe the holographic bulk geometry. Assuming an emergent locality in the holographic bulk, the expectation is that the bulk effective theory $S_\text{eff}[\zeta]$ will take the following form in the continuum limit,
\eq{\label{eq:Seff}S_\text{eff}[\zeta]=\int_\scM g^{\mu\nu}\partial_\mu\zeta^*\partial_\nu\zeta+m^2|\zeta|^2+u|\zeta|^4+\cdots,}
which describes the bulk field $\zeta$ on a curved spacetime background $\scM$ equipped with the metric tensor $g^{\mu\nu}$. Strictly speaking, $\zeta$ is not a single field but contains a tower of fields corresponding to different primary operators in the CFT. We choose to focus on the lightest component and model it by a scalar field, as it will dominate the bulk mutual information at large scale. Because the bulk field excitation is massive and can not propagate far, we expect the mutual information between the bulk variables at two different points to decay exponentially with their geodesic distance in the bulk. Following this idea, suppose $\zeta_i=\zeta(x_i,z_i)$ and $\zeta_j=\zeta(x_j,z_j)$ are two bulk field variables, then their distance $d(\zeta_i:\zeta_j)$ can be inferred from their mutual information $I(\zeta_i:\zeta_j)$ as follows
\eq{\label{eq:d}d(\zeta_i:\zeta_j)=-\xi \ln \frac{I(\zeta_i:\zeta_j)}{I_0},} 
where the correlation length $\xi$ and the information unit $I_0$ are global fitting parameters. 

To estimate the mutual information among bulk field variables, we take a quadratic approximation of the bulk effective action $S_\text{eff}[\zeta]\simeq \sum_{ij}\zeta_i^*K_{ij}\zeta_j=\zeta^\dagger K \zeta$, ignoring the higher order interactions of $\zeta$ for now. This amounts to relaxing the prior distribution of the bulk field $\zeta$ to a correlated Gaussian distribution
\eq{\label{eq:Pblk}P'_\text{prior}[\zeta]=\frac{1}{\sqrt{\det(2\pi K^{-1})}}e^{-\zeta^\dagger K \zeta}.}
The kernel matrix $K$ is carefully designed to ensure positivity and bulk locality (see \appref{sec:kernel} for more details). To determine the best fit of $K$, we initiate the stage II training to learn the prior distribution with the EHM fixed at its optimal solution obtained in the stage I training, as illustrated in \figref{fig:network}(c). We use the reparametrization trick~\cite{Kingma2013AVB} to sample the bulk field $\zeta$ from the correlated Gaussian in \eqnref{eq:Pblk}, then $\zeta$ is pushed to the boundary by the fixed EHM to evaluate the loss function $\scL$ in \eqnref{eq:L}, and the gradient signal can back-propagate to train the kernel $K$. As we relax the Gaussian kernel $K$ for training, we can see that the loss function will continue to drop in the stage II, as shown in \figref{fig:network}(d). This indicates that the Gaussian model is learning to capture the residual bulk field correlation (at least partially), such that the overall performance of generation gets improved. One may wonder why not training the generative model $G$ and bulk field distribution $P_\text{prior}[\zeta]$ jointly. This is because there is a trade-off between these two objectives. For example, one can weaken the disentanglers in $G$ and push more correlation to the bulk field distribution $P_\text{prior}[\zeta]$. Such trade-off will undermine our objective of minimizing bulk mutual information in training a good EHM, therefore the two training stages should be separated, or at least assigned very different learning rates. Intuitively, the machine learns the background geometry in the stage I training and the bulk field theory (to the quadratic order) in the stage II training. The trade-off between the two training stages resembles the interchangeable roles between matter and spacetime geometry in a gravity theory.

\begin{figure}[htbp]
\begin{center}
\includegraphics[width=\columnwidth]{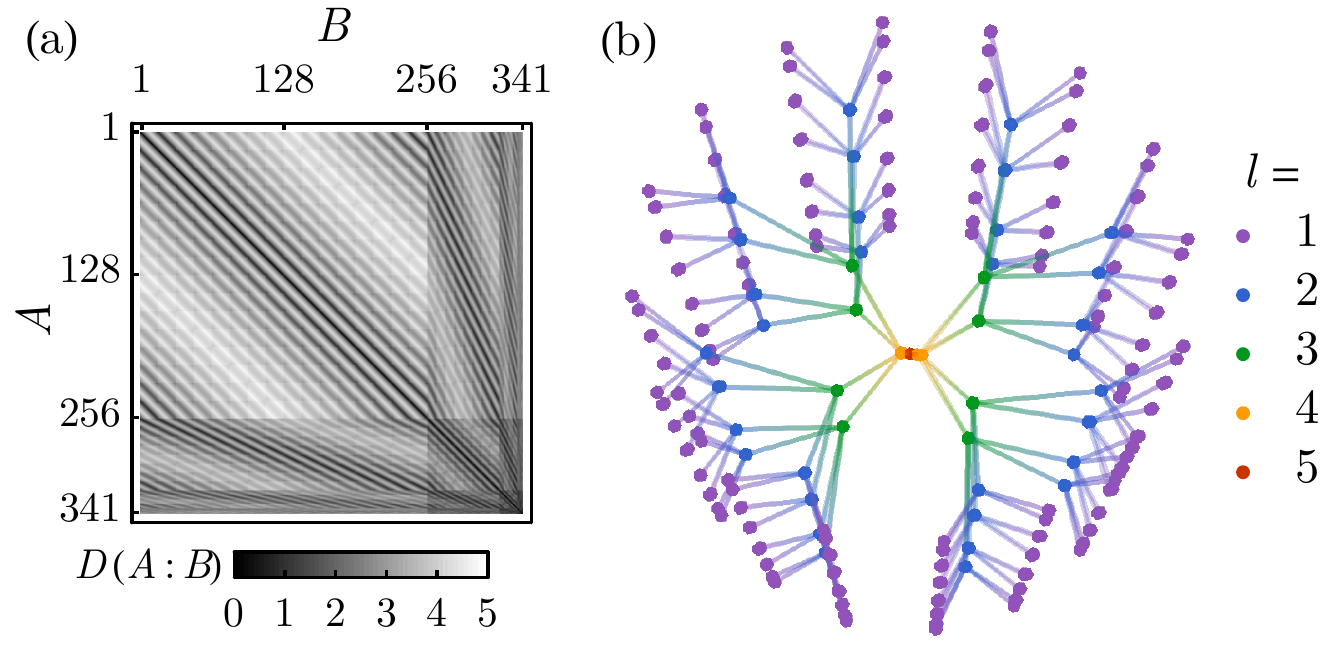}
\caption{(a) Distance matrix $D(A:B)$, indexed by the decimator indices $A,B$, obtained based on \eqnref{eq:D}. (b) Visualization of the bulk geometry by multidimensional scaling projected to the leading three principle dimensions. Each point represent a decimator in the neural network, colored according to layers from UV to IR. The neighboring UV-IR links are add to guide the eye.}
\label{fig:geometry}
\end{center}
\end{figure}

\begin{figure}[htbp]
    \centering
    \includegraphics[width = 0.95\linewidth]{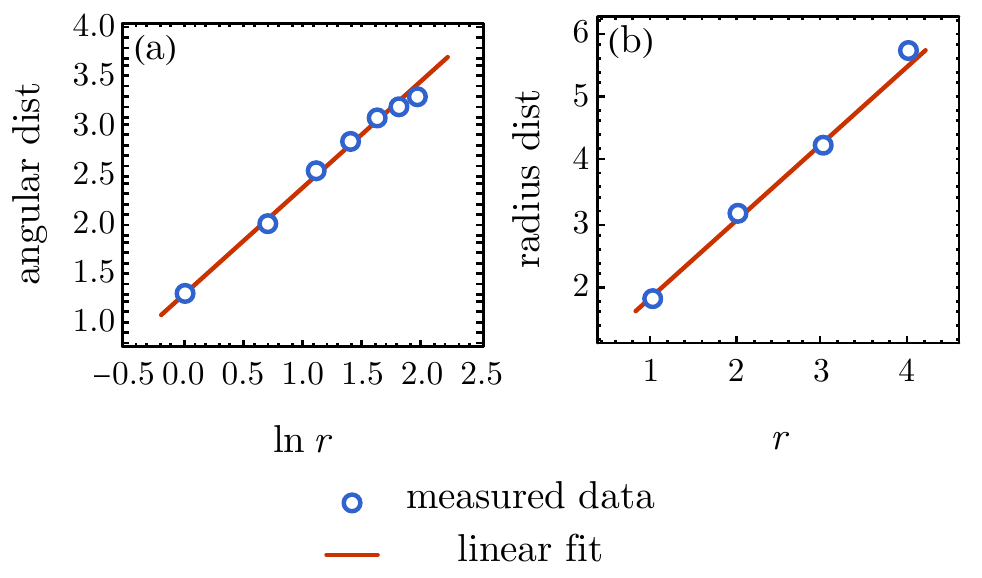}
    \caption{Distance scaling along (a) the radius and (b) the angular direction.\label{fig:distance}}
\end{figure}

After the stage II training, we obtain the fitted kernel matrix $K$. The mutual information $I(\zeta_i:\zeta_j)$ can be evaluated from
\eq{I(\zeta_i:\zeta_j)=-\frac{1}{2}\ln\Big(1-\frac{\langle\zeta_i^*\zeta_j\rangle}{\langle\zeta_i^*\zeta_i\rangle\langle\zeta_j^*\zeta_j\rangle}\Big),}
where the bulk correlation $\langle\zeta_i^*\zeta_j\rangle =(K^{-1})_{ij}$ is simply given by the inverse of the kernel matrix $K$. Then we can measure the holographic distance $d(\zeta_i:\zeta_j)$ between any pair of bulk variables $\zeta_{i}$ and $\zeta_j$ following \eqnref{eq:d}. To probe the bulk geometry, we further define the distance between two decimators $A$ and $B$ to be the average distance between all pairs of bulk variables separately associated to them,
\eq{D(A:B)=\mathop{\mathrm{avg}}\limits_{\zeta_i\in A,\zeta_j \in B}d(\zeta_i:\zeta_j).\label{eq:D}}
The result is presented in \figref{fig:geometry}(a). To visualize the bulk geometry qualitatively, we perform a multidimensional scaling to obtain a three-dimensional embedding of the decimators in \figref{fig:geometry}(b). One can see a hyperbolic geometry emerges in the bulk. To be more quantitative, we label each decimator by three coordinates $(x^1,x^2,z)$, where $x=(x^1,x^2)$ denotes its center position projected to the holographic boundary and $z=2^l$ is related to its layer depth $l$ (ascending from UV to IR). We found that the measured distance function follows the scaling behavior
\eqs{D(x^1,x^2,z:x^1+r,x^2,z)&\propto\ln r,\\
D(x^1,x^2,z:x^1,x^2,z+r)&\propto r,} 
as demonstrated in \figref{fig:distance}. These scaling behaviors agree with the geometry of a three-dimensional hyperbolic space $H^3$, which corresponds to the AdS$_3$ spacetime under the Wick rotation of the time dimension. This indicates that the emergent bulk geometry is indeed hyperbolic at the classical level.

Our result demonstrates that the Luttinger liquid CFT can be approximately dual to a massive scalar fields on AdS$_3$ background geometry. The duality is only approximate because we have assumed a classical geometry in the bulk, ignoring all the gravitational fluctuations. In AdS$_3$/CFT$_2$ correspondence, the bulk gravitational coupling $G_N=3\ell/2c$ is inversely proportional to the central charge $c$ of the CFT.\cite{Brown1986CCCRASEFTG} The
Luttinger liquid CFT has a relatively small central charge $c=1$ and hence a large gravitational coupling in the bulk, so we should not expect a classical dual description. It would be more appropriate to consider holographic CFTs which admit classical duals. However, our current method only applies to lattice field theories of bosons with explicit action functionals, which prevent us to study interesting holographic CFTs. Generalizing the neural RG approach to involve fermions and gauge fields and to work with continuous spacetime will be important directions for future development.

\section{Summary and Discussions}
In conclusion, we introduced the neural RG algorithm to allow automated construction of EHM by machine learning instead of human design. Previously, the EHM was only designed for free fermion CFT. Using machine learning approaches, we are able to develop more general EHMs that also apply to interacting field theories. Given the QFT action as input, the machine effective digests the information contained in the action and encode it into the structure of the EHM network, which represents the emergent holographic geometry. Our result provides a concrete example that the holographic spacetime geometry can emerge as the optimal generative network of a quantum field theory.\cite{Dong2018SOGNQSRQ} The obtained EHM simultaneously provides an information-preserving RG scheme and a generative model to reproduce the QFT, which could be useful for both inference and sampling tasks. 

However, as a version of EHM, our approach also bares the limitations of EHM. By construction, the bulk geometry is discrete and classical, such that the model can not resolve the sub-AdS geometry and can not capture gravitational fluctuations. Recent development of neural ordinary differential equation approaches\cite{Chen2018NODE,Zhang2018MFGM,Grathwohl2018FFCDSRGM} are natural ways to extend our flow-based generative model to the continuum limit. Continuous formulation of real-space RG has been discussed in the context of gradient flows\cite{Fujikawa2016GFET,Abe2018GFRG,Carosso2018NRONSUGF} and trivializing maps\cite{Luscher2010TMWFA}, where the RG flow equations are human-designed. Our research may pave way for machine-learned RG flow equations for continuous holographic mappings. Our formalism also allows the inclusion of gravitational fluctuations in principle, by relaxing optimization to allow superposition of different EHMs. Our analysis indicates that the fluctuation of neural network parameters is related to the bulk gravitational fluctuation. The machine-learned EHM provides us a starting point to investigate the corrections on top of the classical geometry approximation, which may enable us to go beyond holographic CFTs and study the quantum gravity dual of generic QFTs. Another feature of EHM is that it is a one-to-one mapping of field configurations (operators) between bulk and boundary, while in holographic duality, a local bulk operator can be mapped to multiple boundary operators in different regions. A resolution\cite{Almheiri2015BLQECA,Pastawski2015HQECMBC} of the paradox is that the non-unique bulk-boundary correspondence only applies to the low-energy freedoms in the bulk, which can be encoded on the boundary in a redundant and error-correcting manner. The bidirectional holographic code (BHC)\cite{Yang2016BHCSL,Qi2017HCSFRTN} was proposed as an extension of the EHM to capture the error-correction property of the holographic mapping. Extending our current network design to realize machine-learned BHC will be another open question for future research.

\begin{acknowledgments}
We acknowledge the stimulating discussions with Xiao-Liang Qi, John McGreevy, Maciej Koch-Janusz, C\'edric B\'eny, Koji Hashimoto, Wenbo Fu and Shang Liu.
S.H.L and L.W. are supported by the National Natural Science Foundation of China under the Grant No.~11774398 and the Strategic Priority Research Program of Chinese Academy of Sciences Grant No.~XDB28000000.
\end{acknowledgments}

\bibliography{ML}
\bibliographystyle{apsrev}
\onecolumngrid
%\vspace{24pt}
%\twocolumngrid
\newpage
\appendix
\section{Minimal Bulk Mutual Information Principle}\label{sec:minBMI}
The maximal real-space mutual information (maxRMI) principle proposed in \refcite{Koch-Janusz2018MINNRG,Lenggenhager2018ORGTFIT} aims to maximize the mutual information between the coarse-grained field and the fine-grained field in the surrounding environment at a single RG step. In this section, we show that the maxRMI principle can be derived from our minimal bulk mutual information (minBMI) principle under certain assumptions.

\begin{figure}[H]
    \centering
    \includegraphics[width = 0.25\linewidth]{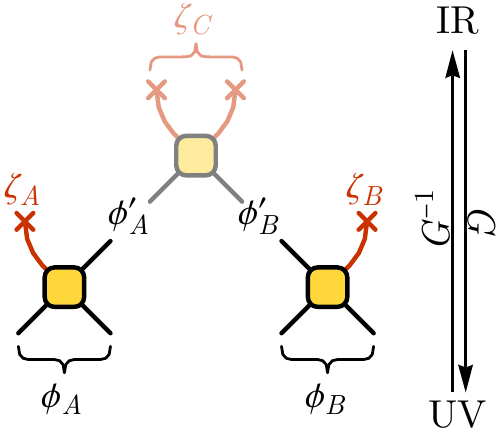}
    \caption{Functional dependence of variables in the neural-RG network. Each block represents a bijective map.}
    \label{fig:MI}
\end{figure}

Let us set up the problem based on Fig.\ref{fig:MI}. Assuming $\phi_A$ and $\phi_B$ are field configurations in two neighboring regions $A$ and $B$ in the UV layer. Under one step of the RG transformation, $\phi_A$ gets mapped to the coarse grained variable $\phi'_A$ and the bulk variable $\zeta_A$, and the mapping is bijective. Similarly, another bijection takes $\phi_B$ to $\phi'_B$ and $\zeta_B$. Eventually, $\phi'_A$ and $\phi'_B$ will be mapped to the bulk field $\zeta_C$ in deeper IR layers. Therefore the random variables appeared in \figref{fig:MI} are related by the following bijections $f_{A}$, $f_B$, $f_C$ as
\eq{(\phi'_A,\zeta_A)=f_A(\phi_A), (\phi'_B,\zeta_B)=f_B(\phi_B), \zeta_C=f_C(\phi'_A,\phi'_B).}
What are the information theoretical principles to guide the bijections $f_A$, $f_B$, $f_C$ toward good RG transformations? We propose the minBMI principle that these bijections should minimize the mutual information among the bulk variables, 
\eq{\label{eq:min}\min I(\zeta_A:\zeta_B)+ I(\zeta_A:\zeta_C)+ I(\zeta_B:\zeta_C).}
\refcite{Koch-Janusz2018MINNRG,Lenggenhager2018ORGTFIT} propose another principle, the maxRMI principle, that the RG transformation should maximize the mutual information between the coarse grained variable (such as $\phi'_A$) and its environments (such as $\phi_B$),
\eq{\label{eq:max}\max I(\phi'_A:\phi_B).}
We can show that the objective of the maxRMI in \eqnref{eq:max} is consistent with the objective of the minBMI in \eqnref{eq:min} in the limit of UV-IR decoupling.

The minBMI principle aims to minimize mutual information among all bulk variables, both between different RG scales and within the same RG scale. Its objective has a broader scope than the maxRMI principle, because the later does not specify its objectives across the RG scales. So to make a connection between these two principles, one must first restrict the scope of the minBMI principle to a single layer. This can be achieved by assuming that there is no mutual information between bulk variables at different RG scales. In our setup, this corresponds to $I(\zeta_A,\zeta_B:\zeta_C)=0$, which factorizes the joint probability $p(\zeta_A,\zeta_B,\zeta_C)=p(\zeta_A,\zeta_B)p(\zeta_C)$ and decouples the bulk variables between UV and IR. As a result, the mutual information between any bulk variables across different RG scales vanishes $I(\zeta_A:\zeta_C)=I(\zeta_B:\zeta_C)=0$. This already minimizes the bulk mutual information across layers and reduces the minBMI objective in \eqnref{eq:min} to
\eq{\min I(\zeta_A:\zeta_B).}
In this UV-IR decoupled limit, we can prove that $\max I(\phi'_A:\phi_B)$ and $\min I(\zeta_A:\zeta_B)$ are equivalent.

The proof starts by considering the mutual information between $\phi_A$ and $\phi_B$. We can see that
\eqs{\label{eq:MIproof}I(\phi_A:\phi_B)&=I(\phi'_A,\zeta_A:\phi_B)\\
&=I(\phi'_A:\phi_B)+I(\zeta_A:\phi_B)\\
&=I(\phi'_A:\phi_B)+I(\zeta_A:\phi'_B,\zeta_B)\\
&=I(\phi'_A:\phi_B)+I(\zeta_A:\phi'_B)+I(\zeta_A:\zeta_B)\\
&=I(\phi'_A:\phi_B)+I(\zeta_A:\zeta_B).\\}
Here we have used the bijective property of $f_A$, $f_B$, $f_C$ to obtain $I(\phi_A:\phi_B)=I(\phi'_A,\zeta_A:\phi_B)$, $I(\zeta_A:\phi_B)=I(\zeta_A:\phi'_B,\zeta_B)$ and $I(\zeta_A:\zeta_C)=I(\zeta_A:\phi'_A,\phi'_B)$. In the UV-IR decoupled limit, $I(\zeta_A:\zeta_C)=0$, so $I(\zeta_A:\phi'_A,\phi'_B)=0$, which further implies $I(\zeta_A:\phi'_A)=I(\zeta_A:\phi'_B)=0$. With these relations, all steps in \eqnref{eq:MIproof} are justified. On the left hand side, $I(\phi_A:\phi_B)$ is determined by the field theory in the UV layer, which can be treated as a constant. For the given amount of information between regions $A$ and $B$, \eqnref{eq:MIproof} tells us that $I(\phi'_A:\phi_B)$ and $I(\zeta_A:\zeta_B)$ are competing for information resources. Therefore maximizing $I(\phi'_A:\phi_B)$ is equivalent to minimizing $I(\zeta_A:\zeta_B)$.

We can apply this argument layer by layer. Then to achieve the objective of the maxRMI principle, we need to minimize mutual information among bulk variables in the same RG scale, which is precisely the statement of the minBMI principle when restricted to each layer. In this sense, the maxRMI and minBMI principles are consistent. However, the minBMI principle actually relaxes the assumption that bulk variables at different RG scales are fully decoupled. Instead, we want to minimize mutual information among all bulk variables, including those across the scales. In this sense, the minBMI principle is more general than the maxRMI principle.

\section{Design of Bijectors}\label{sec:bijectors}
We designed a set of symmetry-persevere bijectors to making sure that $U(1)$ symmetry of the boundary is preserved at each bijector. For the generative process, at each RG step, it takes the four complex degrees of freedom and they go through three layers of bijectors: $\mathcal{S}$, $\mathcal{O}$, and $\mathcal{A}$.

\textbf{I. Scaling layer($\mathcal{S}$):} At scaling layer, each complex variables $\phi_{i}$ is multiplied by a factor $e^{\lambda_i}$. The inverse and the Jacobian of this transformation can be obtained easily.

\textbf{II. Orthogonal transformation layer($\mathcal{O}$):} The orthogonal transformation in disentangler and decimator is in general an $O(4)$ transformation. In stead, we implemented it by stacking multiple $O(2)$ transformations. In \figref{appendix:O_transform}(a), each blue block represents the matrix:
\eq{
M_{\text{blue}}(\theta_{i})=\begin{pmatrix}
    \sin \theta_{i} & \cos \theta_{i}\\
    \cos \theta_{i} & -\sin \theta_{i}
\end{pmatrix},
}
and the orange block in \figref{appendix:O_transform}(b) represents the matrix:
\eq{
M_{\text{orange}}(\theta_{i})=\begin{pmatrix}
    \cos \theta_{i} & -\sin \theta_{i}\\
    \sin \theta_{i} & \cos \theta_{i}
\end{pmatrix}.
}
$\theta_{i}$ in those blocks are training parameters. The arrangement of the type I and type II blocks are such designed that when $M_{\text{blue}}(\theta_{i}=\pi/4)$ and $M_{\text{orange}}(\theta_{i}=0)$ the network reproduces the ideal EHM originally proposed in \refcite{Qi2013EHMESG}. We initialize the parameter to this ideal limit. 

\begin{figure}[H]
    \centering
    \includegraphics[width = 0.45\linewidth]{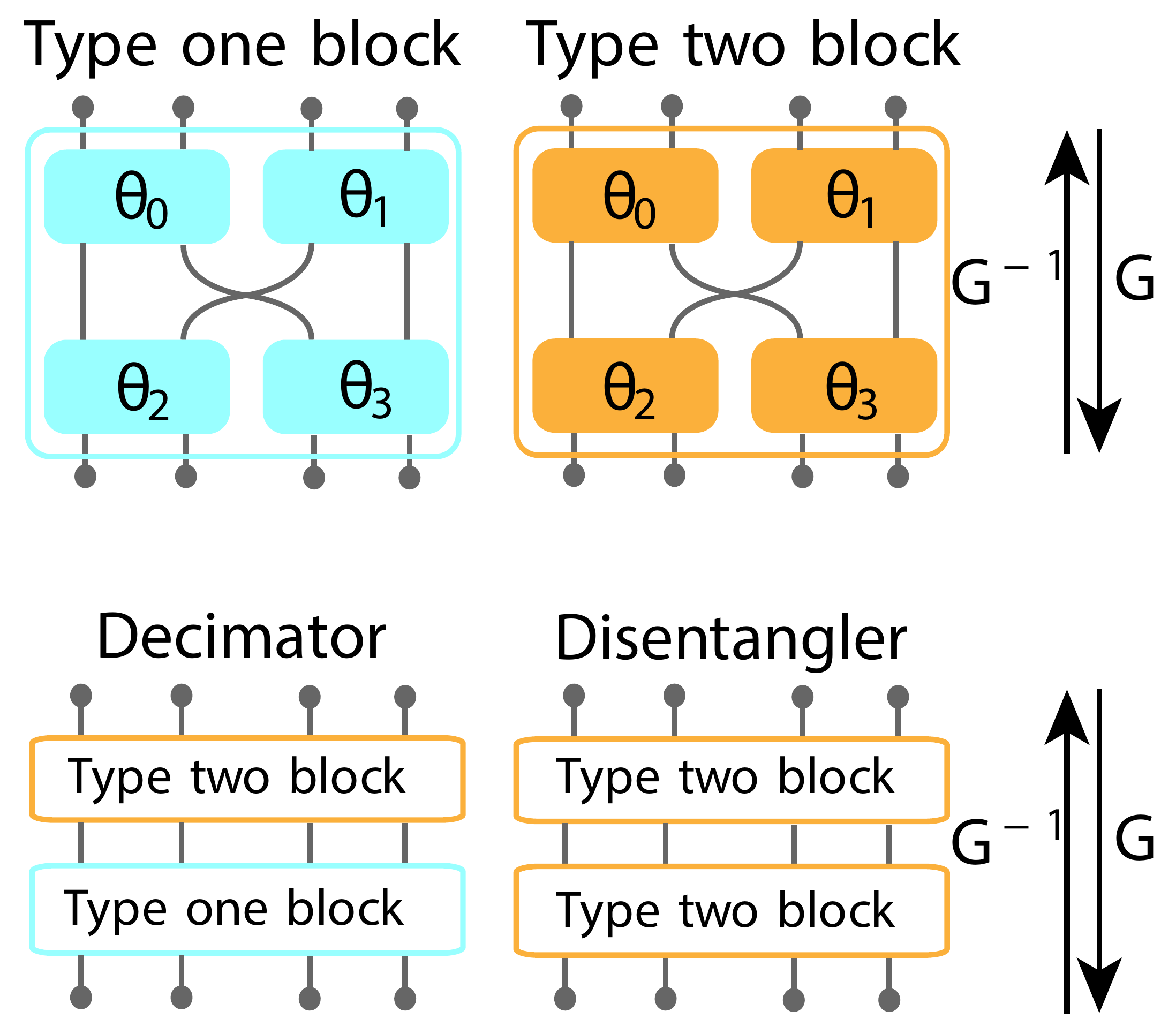}
    \caption{\label{appendix:O_transform}Orthogonal transformation.}
\end{figure}
\textbf{III. Non-linear layer($\mathcal{A}$)} For non-linear part, we use the amplitude hyperbolic functions for complex field $\phi_{i}$. In coarse-graining direction, it acts in the following,
\begin{equation}
  \begin{array}{l} 
  \Re(\zeta) = \sinh{|\phi|} \dfrac{\Re \phi}{|\phi|}; \\
  \Im(\zeta) = \sinh{|\phi|} \dfrac{\Im \phi}{|\phi|}.
  \end{array}
\label{eq:AmpHForward}
\end{equation}
The corresponding inverse and Jacobian can be calculated easily.

\section{Neural Network Training}\label{sec:NNtraining}
All the training parameters of our neural RG network are contained in scaling bijectors and orthogonal transformation bijectors as illustrated in Appendix \ref{sec:bijectors}. We imposed translation invariance of our network at each layer, due to translation invariance of the system at each energy scale. The total number of training parameters scale with $O(\log(N))$, where $N$ is the size of boundary theory. The prefactor depends on the depth of bijector neural networks. In our case, the total number of training parameters are $24\log_2 N$. In order for faster convergence of the training, we first set learning rate for parameters contained in scaling bijectors as $10^{-2}$, and gradually reduce it to $10^{-4}$. And the learning rate for parameters contained in orthogonal transformation bijectors is always $10^{-4}$. 

\section{Monte Carlo Sampling Efficiency}\label{sec:MC}
\begin{figure}[H]
    \centering
    \includegraphics[width = 0.45\linewidth]{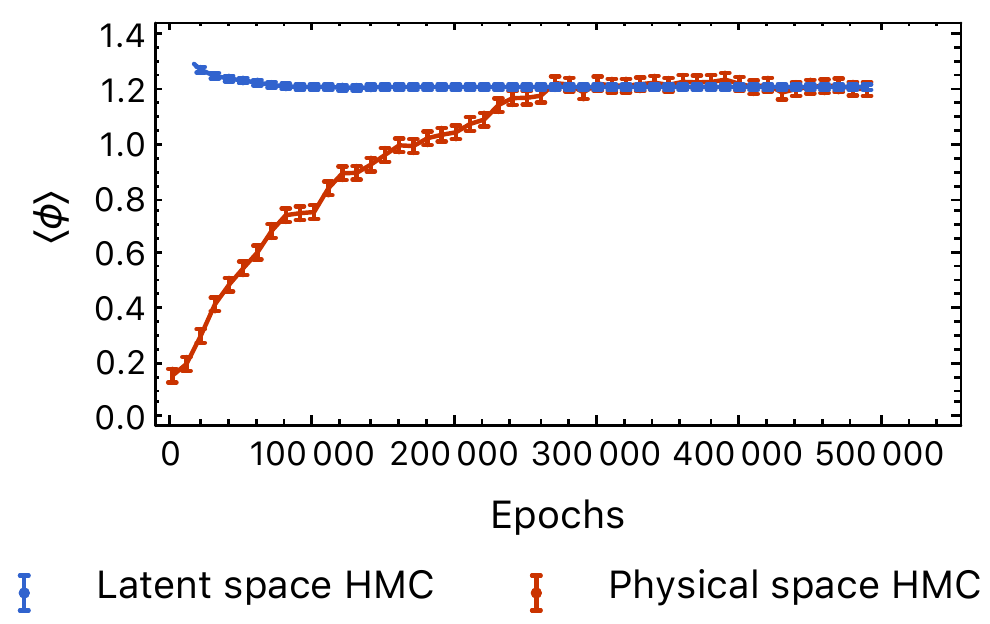}
    \caption{\label{appendix:mcmc_conv}MCMC result.}
\end{figure}
We tested numerical efficiency of our method by comparing convergence rate between Hamiltonian Monte Carlo(HMC) on the boundary system, and HMC in the bulk system. Both method are implemented using TensorFlow probability API with same parameters. The result is shown in Fig.\ref{appendix:mcmc_conv}. As we can see, the HMC in the bulk system converges faster than the HMC on the boundary system.

\section{Design of the Correlated Gaussian Prior}\label{sec:kernel}
In finding the effective bulk field theory, we assume the bulk field is very massive. Under this assumption, higher-order interaction terms are irrelevant. Therefore, we use a correlated Gaussian distribution with positive definite kernel matrix $K$ as our effective bulk field theory. We also assumed locality of our effective bulk field theory, which means $K_{ij}$ is non-zero if and only if $\zeta_{i}$ and $\zeta_{j}$ are nearest neighbors in the bulk, including neighbors inter-scale and intra-scale. To further reduce the fitting parameters of matrix $K$, we also imposed translation invariance of the bulk field at each scale. This is reasonable, because our RG scheme also has the translation invariance at each scale. 

To ensure matrix $K$ is positive definite, we decomposed matrix $K$ into a set of positive semi-definite matrix and a mass term. Particularly, 
\eq{K = \sum_{\langle ij\rangle}\lambda_{ij}(|i\rangle\langle i|+|j\rangle \langle j|-|i\rangle \langle j|-|j\rangle\langle i|)+m\mathbb{I},}
where $\mathbb{I}$ is the identity matrix, and $\lambda_{ij}$ and $m$ are positive numbers. This ensures matrix $K$ we constructed is positive definite.

%\section{Hyperbolic Space}
%\begin{figure}[H]
%    \centering
%    \includegraphics[width = 0.35\linewidth]{appendix_hyperbolic_space}
%    \caption{\label{appendix:geo_dis}Geodesic distance in hyperbolic space.}
%\end{figure}
%The metric of Euclidean AdS$_{d+1}$ space (the hyperbolic space) is given by
%\eq{\dd s^{2}=g_{\mu \nu}\dd z^{\mu}\dd z^{\nu}=\dfrac{1}{z_{0}^{2}}(\dd z_{0}\dd z_{0}+\dd \textbf{z}\cdot \dd \textbf{z})
%}
%The geodesic equation is given by
%\eqs{
%\dfrac{\dd^{2}z_{0}}{\dd \tau}-\dfrac{1}{z_{0}}\left( \dfrac{\dd z_{0}}{\dd \tau}\right)^{2}+\dfrac{1}{z_{0}}\left(\dfrac{\dd z}{\dd \tau} \right)^{2}=0,\\
%\dfrac{\dd^{2} \textbf{z}}{\dd \tau}-\dfrac{2}{z_{0}}\dfrac{\dd z_{0}}{\dd \tau}\dfrac{\dd \textbf{z}}{\dd \tau}=0.
%}
%As shown in \figref{appendix:geo_dis}, given any two points $z$ and $w$ in hyperbolic space, there is a unique semi-circle connecting them. The geodesic distance can be calculated as 
%\eq{
%d(z|w)=\text{arccosh} \dfrac{z_{0}^{2}+w_{0}^{2}+(\textbf{z}-\textbf{w})^{2}}{2z_{0}w_{0}}.\label{eq:geo_dis}
%}
%

%\begin{figure}[H]
%    \centering
%    \includegraphics[width = 0.35\linewidth]{triangle_inequality}
%\end{figure}

%\begin{thebibliography}{9}
%\bibitem{NeuralRG}Shuo-Hui Li, and Lei Wang. Phys. Rev. Lett. 121, 260601(2018)
%\bibitem{xiaoliang}X.-L. Qi, arXiv preprint arXiv:1309.6282 (2013).
%\bibitem{xiaogang}Quantum field theory of many-body systems. Xiao-Gang Wen.
%\end{thebibliography}

\end{document}